\documentclass[twocolumn,twocolappendix]{aastex6}
\usepackage{epstopdf}
\usepackage{graphicx}
\usepackage{natbib}
\usepackage{float}
\usepackage[utf8]{inputenc}
\usepackage{amsmath}
\usepackage{bm}

\begin{document}

\title{Computing the Polarimetric and Photometric Variability of Be Stars}
\author{K. C. Marr\altaffilmark{1}, C. E. Jones\altaffilmark{1} and R. J. Halonen\altaffilmark{2}}
\altaffiltext{1}{Department of Physics and Astronomy, Western University, London, ON, N6A 3K7, Canada}
\altaffiltext{2}{School of Arts \& Science, Red Deer College, Red Deer, AB, T4N 5H5, Canada}

\begin{abstract}
We investigate variations in the linear polarization as well as in the V-band and B-band colour-magnitudes for classical Be star disks. We present two models: disks with enhanced disk density and disks that are tilted or warped from the stellar equatorial plane. In both cases, we predict variation in observable properties of the system as the disk rotates. We use a non-LTE radiative transfer code \textsc{bedisk} (Sigut \& Jones) in combination with a Monte Carlo routine that includes multiple scattering (Halonen et al.) to model classical Be star systems. We find that a disk with an enhanced density region that is one order of magnitude denser than the disk's base density shows as much as ${\sim}0.2\%$ variability in the polarization while the polarization position angle varies by ${\sim}8^{\circ}$. The $\Delta$V magnitude for the same system shows variations of up to ${\sim}0.4$ magnitude while the $\Delta$(B-V) colour varies by at most ${\sim}0.01$ magnitude. We find that disks tilted from the equatorial plane at small angles of ${\sim}30^{\circ}$ more strongly reflect the values of polarization and colour-magnitudes reported in the literature than disks tilted at larger angles. For this model, the linear polarization varies by ${\sim}0.3\%$, the polarization position angle varies by ${\sim}60^{\circ}$, the $\Delta$V magnitude varies up to 0.35 magnitude, and the $\Delta$(B-V) colour varies up to 0.1 magnitude. We find that the enhanced disk density models show ranges of polarization and colour-magnitudes that are commensurate with what is reported in the literature for all sizes of the density enhanced regions. From this, we cannot determine any preference for small or large density enhanced regions.

\end{abstract}


\section{Introduction} \label{sec:intro}

Classical Be (B-emission) stars are main sequence, B-type stars for which there is observational evidence of the presence of a geometrically flat decretion disk of ionized gas in Keplerian motion \citep{rivinius2013}. The evolution of the circumstellar disks of Be stars is thought to be facilitated by viscosity, as described by the viscous disk decretion (VDD) model \citep{lee1991}. The rapid rotation of these massive stars \citep{slettebak1982} leads to the formation of the disk, which is sustained through mass ejection into the disk, in discrete outburst events \citep{rivinius1998, kee2014} or possibly continuously over long periods (e.g. \citet{caballero2016}). The defining characteristics of classical Be stars include spectral line emission in the Balmer series, periodic brightness and line profile variations over time scales ranging from minutes to decades \citep{harmanec1983}, an excess of IR and radio continuum emission, and linearly polarized light \citep{hall1950, behr1957}.

The linear polarization signature of Be stars can vary over time and can account for up to ${\sim}2\%$ of the total light emitted. The source of the polarization is electron scattering of the star's light as it interacts with the disk. Through electron scattering, the light within the disk becomes linearly polarized perpendicular to the plane containing the incident and scattered radiation \citep{quirrenbach1997}. The polarized light from a Be star provides a wealth of information on the geometry and physical nature (e.g. chemical composition, ionization levels, opacity) of the disk without the need of resolving the source. The strength of the polarization signature is proportional to the number of particles available to scatter the light. However, there is a complex interaction between the scattering and absorption processes in the disk, which impart a wavelength dependence on the polarization signature. The hydrogen bound-free absorption imprints a saw-tooth shape on the polarization as a function of wavelength that weakens as wavelength increases and rapidly increases at each series limit (see Figure 1 of \citet{halonen2013b}).

Be stars exhibit variability on many different timescales ranging from hours to decades, a characteristic which suggests the involvement of various astrophysical phenomena. The fastest variations in Be stars, called ultra-rapid variations \citep{huang1986}, occur on timescales of hours and are associated with $\beta$-Cephei type pulsations if recurring \citep{balona1991}. The short-term variations occurring over periods of days are primarily driven by non-radial pulsations \citep{baade1982, rivinius2003}, but stellar rotation \citep{balona1990, balona1995} and localized mass ejections in the inner disk may produce variations of similar length. The intermediate length variations occur over weeks to decades. They are characterized by variability in emission-line strength, often measured by the V/R (violet to red) ratio of doubly-peaked lines \citep{hanuschik1996, rivinius2006}, and are due to changes in the disk as a result of binarity on shorter terms, or the presence of precessing one-arm density waves on longer terms. The longest timescale for variations measured in Be stars occur over years to decades. These variations are classified by the loss and return of spectral emission properties and have been linked to the growth and dissipation of the circumstellar disk, for example $\pi$ Aqr's (HD 212571) disk dissipation between 1986 and 1996 \citep{wisniewski2010, draper2011}.

\citet{huang1986} suggested that there may be correlation between the short and long-term variations in Be stars. The local mass ejection of stellar surface material into the disk is theorized to cause an appreciable increase in density of local regions which may lead to variability in the observational characteristics of the stars \citep{okazaki1991}. Since this prediction, there have been many studies of one-armed density waves. Notably, \citet{carciofi2009} were able to confirm the presence of a spiral density arm in the disk of the Be star $\zeta$ Tauri (HD 37202). Additionally, they were able to show that the oscillation mode must extend to the outer regions of the disk to properly fit the large amplitude of the V/R variations. This suggests that long period variations could result from mass ejection into the disk. However, \citet{carciofi2009}'s model also exhibited inconsistencies with particular observations, such as large variation in the polarization signature over the V/R cycle, which suggests that the innermost disk may not contain the expected density structure.

Perturbations in the circumstellar disk of Be star systems may also cause long period variations as seen, for example, in the behaviour of $\gamma$ Cas (HD 5394) and 59 Cyg (HD 200120). These two stars are notable as having showed two successive shell events at the same time as a variation in emission-line width. According to \citet{hummel1998}, $\gamma$ Cas varied as such from 1934 to 1940, after which its disk dissipated, while the same variations occurred for 59 Cyg from June 1973 to April 1975. In the same work, he suggested that these variations are the result of a disk that has been tilted from the equatorial plane of the star. Such a system could form as between one third and half of Be stars are found to be in binary systems, where there is a reasonable chance that the companion star would be misaligned with the equatorial plane of the Be star. Through tidal interactions, the the disk can become inclined with respect to the star, and align with the companion's orbital plane. See \citet{martin2011} for numerical models involving tidal warping and the evolution of misaligned disks. The variation in emission-line width and profile can be explained by the change in geometry of the system on the plane of the sky as the disk rotates while in a tilted state.

In this study, we investigate two cases of Be star disk perturbations, density enhancement (DE) and tilted disk (TD), by studying the linear polarization and the B-band and V-band magnitudes predicted by models. Through the use of a non-LTE radiative transfer code called \textsc{bedisk} \citep{sigut2007}, the self-consistent radiative equilibrium models of the disk were acquired. From these models, the Stokes parameters were computed using the Monte Carlo routine \textsc{mctrace} \citep{halonen2013}.  In Section 2, we discuss the computational method used in creating each model. Section 3 provides the results of the Density Enhancement (DE) model while Section 4 gives the results of the Tilted Disk (TD) model.. Finally, in Section 5, we discuss the results of each model, we compare our predictions to the observations in other works, and we summarize the findings of this report.

\section{Computational Modelling} \label{sec:models}

\subsection{\textsc{bedisk}}
In order to determine the Stokes parameters for the perturbed disks of Be stars, \textsc{bedisk} was used to create models of axisymmetric disks. This modelling routine creates a two-dimensional cross-section of the disk with self-consistent thermal and density structures. The disk is axisymmetric around the star's rotational axis and symmetric above and below the equatorial plane. A synthetic spectrum emulates the radiative output of the central star based on the desired model atmosphere from \citet{kurucz1993}, which is selected through the $\log(g)$ and T$_{\rm eff}$ of the star considered. (see \citet{sigut2007} for more detail). Multiple studies with \textsc{bedisk} have shown its utility in modelling spectroscopy \citep{arcos2017, jones2011}, interferometry \citep{jones2009, jones2017}, and also polarimetry and photometry \citep{halonen2013a, halonen2013b}.

For this  work, we assume that the disk density distribution falls off in the equatorial plane by an $R^{-n}$ power law, first suggested by \citet{waters1986} and revisited in \citet{cote1987}, and \citet{waters1987}. The density grid of the modelled disk is expressed in terms of the cylindrical coordinates corresponding to the radial distance from the star $R$, and the height of the disk from the midplane $Z$, and is given by

\begin{equation}
\rho(R,Z) = \rho_{0}\left(\frac{R_{\star}}{R}\right)^{n}e^{{-[Z/H]}^2}.
\end{equation}

Here, $\rho_0$ is the density at the disk's inner edge at $Z=0$, $n$ is the radial density slope, and $H$ is the disk's scale height. At each radial grid point in the disk, the gas was taken to be in approximate vertical hydrostatic equilibrium. In the literature, values of $\rho_0$ are often within the range of $10^{-12}$ gcm$^{-3}$ to $10^{-10}$ gcm$^{-3}$, while those for $n$ are usually between 2 and 4 \citep{rivinius2013}.

The \textsc{bedisk} models for this work were computed using the stellar parameters provided in Table 1.

\begin{table}
\begin{center}
\caption{Reference model stellar parameters}
\label{tab:param1}
\begin{tabular}{lccccc}
\tableline \tableline
Spectral & Radius & Mass & Luminosity & $T_{\rm eff}$ & $\log(g)$\\
Type &($R_{\sun}$)&($M_{\sun}$)&($L_{\sun}$)&(K)& ($\rm cm\, s^{-2}$)\\
\tableline
B2Ve & 5.33 & 9.11 & 4.76 $\times 10^{3}$ & 2.09 $\times 10^{3}$ & 3.9\\
\tableline
\end{tabular}
\tablecomments{Based on values from \citet{cox00}}
\end{center}
\end{table}

\begin{table}
\begin{center}
\caption{Reference model disk parameters}
\label{tab:param2}
\begin{tabular}{lcccc}
\tableline \tableline
Model & Radius & n & Base Density & Enhanced Density\\
 &($R_{\star}$)& &($\rm g\, cm^{-3}$)&($\rm g\, cm^{-3}$)\\
\tableline
DE & 50 & 3.0 & 1.0 $\times 10^{-11}$ & 1.0 $\times 10^{-10}$\\
TD & 50 & 3.0 & 5.0 $\times 10^{-11}$ & -\\
\tableline
\end{tabular}
\end{center}
\end{table}

\subsection{\textsc{mctrace}}

\textsc{mctrace} calculates the Stokes parameters and photometric colours by using the thermal solution and level populations from \textsc{bedisk} as a base model of the circumstellar gas. The routine uses the two-dimensional solution created by \textsc{bedisk} and interpolates it into a three-dimensional grid to simulate the entire disk. It then performs radiative transfer through a Monte Carlo simulation, following the trajectory of each photon packet until it exits the disk after multiple scattering events. The parameters used for each model in \textsc{mctrace} are provided in Table \ref{tab:param2}. For each simulation, $20$ billion photon packets were followed through the disk to ensure appropriate resolution of our results. The routine has been previously used in \citet{halonen2013, halonen2013a, halonen2013b, halonen2015} to synthesize the polarimetric signatures of classical Be stars.

An effective method for quantifying and interpreting polarization with the Stokes parameters was first presented in \citet{stokes1852}. The linear polarization signature can be quantified using the Stokes $Q$ and $U$ parameters. These parameters represent the differences in the total intensity $I$ across two sets of orthogonal axes, with one set rotated $45^{\circ}$ from the other.

Free electrons in the partially ionized disk scatter photons emitted from the disk and the central star. A non-zero polarization signature is observed when the scattered photons are produced in a region that is asymmetric on the plane of the sky of the observer. As a disk becomes less symmetric, fewer photons are polarized along one of the axes, yielding less cancellation of orthogonal polarization vectors and a greater amount of polarization. These asymmetries occur for both non-axisymmetric disks seen at any inclination, and in axisymmetric disks seen at angles other than pole-on. In this work, Section \ref{sec:dem} considers variations in observables which result in non-axisymmetric disks for the DE model, and Section \ref{sec:tdm} looks at variable observables in axisymmetric disks for the TD model.

The relative strength of the polarization can be calculated from the Stokes parameters when added in quadrature. The normalized polarization level are given by $q = Q/I$ and $u = U/I$, and the normalized net linear polarization is then determined by

\begin{equation} \label{eq:pol}
p = (q^{2}+u^{2})^{1/2}.
\end{equation}

The polarization position angle (PA) is given by

\begin{equation} \label{eqn:polposangle}
\theta = \frac{1}{2} \arctan\left(\frac{u}{q}\right).
\end{equation}

As the Stokes parameters change with the inclination of the star, one may expect the strongest polarization signature to occur at $90^{\circ}$. However, in this position the disk attenuates the polarization by re-absorbing much of the scattered light before it exits the disk. As such, the maximum polarization signature is actually seen at an inclination of i $\approx$ $70^{\circ}$ \citep{wood1996, halonen2013a}. In contrast, the weakest polarization signature occurs at $0^{\circ}$ when the disk appears spherically symmetric on the plane of the sky, causing complete cancellation of vibrations from orthogonal directions due to the uniformity of the polarizing planes.

The inclination angle can also affect the photometric variations of a Be star disk, in addition to density enhancements. As shown in Fig. 2 of \citep{rivinius2013}, different wavelengths contribute to the integrated flux of the disk as a function of radial distance from the star. In this work, the V- and B-bands are of primary concern. According to \citep{haubois2012}, 80\% of the V-band flux originates from within 1.8 to 2.5$R_*$ for the same densities considered in the DE model; the B-band is assumed to be similar.

\newpage

\section{Density Enhancement Model} \label{sec:dem}

To study the effects of disk warping, we adopted a disk with a DE region of variable size. First, we used \textsc{bedisk} to create two axisymmetric disks with different densities $\rho_0$ (see Table \ref{tab:param2}). The base density values were chosen to ensure that any effects introduced by the optically thick DE would be apparent in our observables. We note that an increase in the gas density leads to an increase in the scale height of the disk.

\begin{figure}
\centering
\includegraphics[width=\columnwidth]{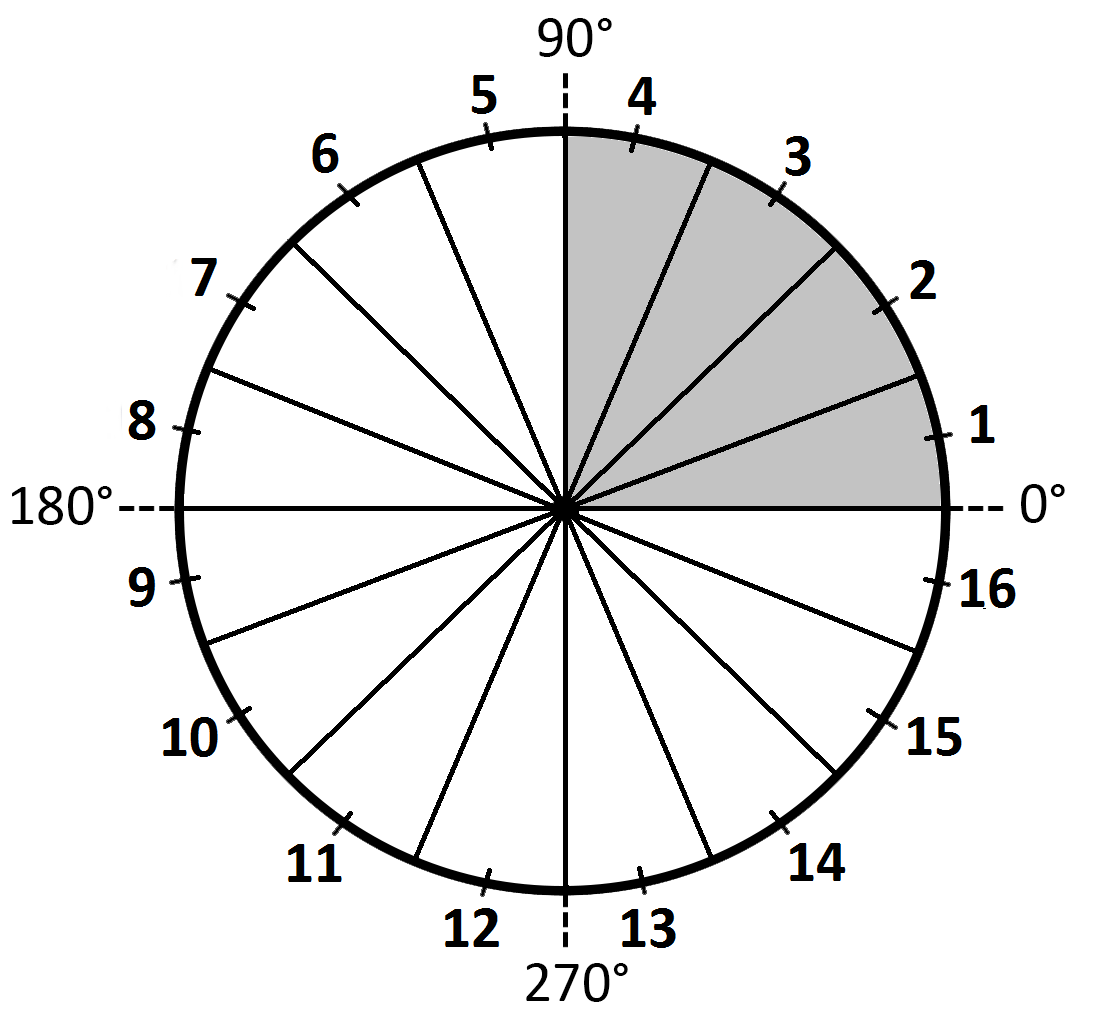}
\caption{In the DE model, the disk is observed at 16 different phases, labelled on this figure, each of which are aligned with the centers of the 16 corresponding disk wedges. An example dense region of the disk, where $\omega = 90^{\circ}$ is shaded in (i.e. first four wedges), which may then be viewed at any of the 16 phases. For simplicity, the central star is not shown.}
\label{fig:outfilepositions}
\end{figure}

These axisymmetric models were then spliced together, with a sector of the greater density model placed within the lower density model (see Figure \ref{fig:outfilepositions}). The spliced model was then evaluated with \textsc{mctrace} to determine the Stokes parameters. \textsc{mctrace} provides 16 different phases at which the disk may be viewed, shown in Figure \ref{fig:outfilepositions}. Plotting the observables sequentially at each phase angle is analogous to viewing the system undergoing a complete rotation in the plane of the sky. For this model, we vary the size of the DE sector, $\omega$, and provide results for sectors with enhanced density of $45^\circ$, $90^\circ$, $180^\circ$ and $270^\circ$.

The observables computed by \textsc{mctrace} are collected in 100 \AA{} wavelength bands centered around pre-specified wavelengths. In this work, the observables were analyzed in the V-band (data averaged from 5250 \AA{} and 5750 \AA), B-band (4450 \AA), and across the Balmer jump from 3800 \AA{} to 3600 \AA.

\newpage

\subsection{Polarization} \label{subsec:dem-pol}

\begin{figure}
\centering
\includegraphics[width=\columnwidth]{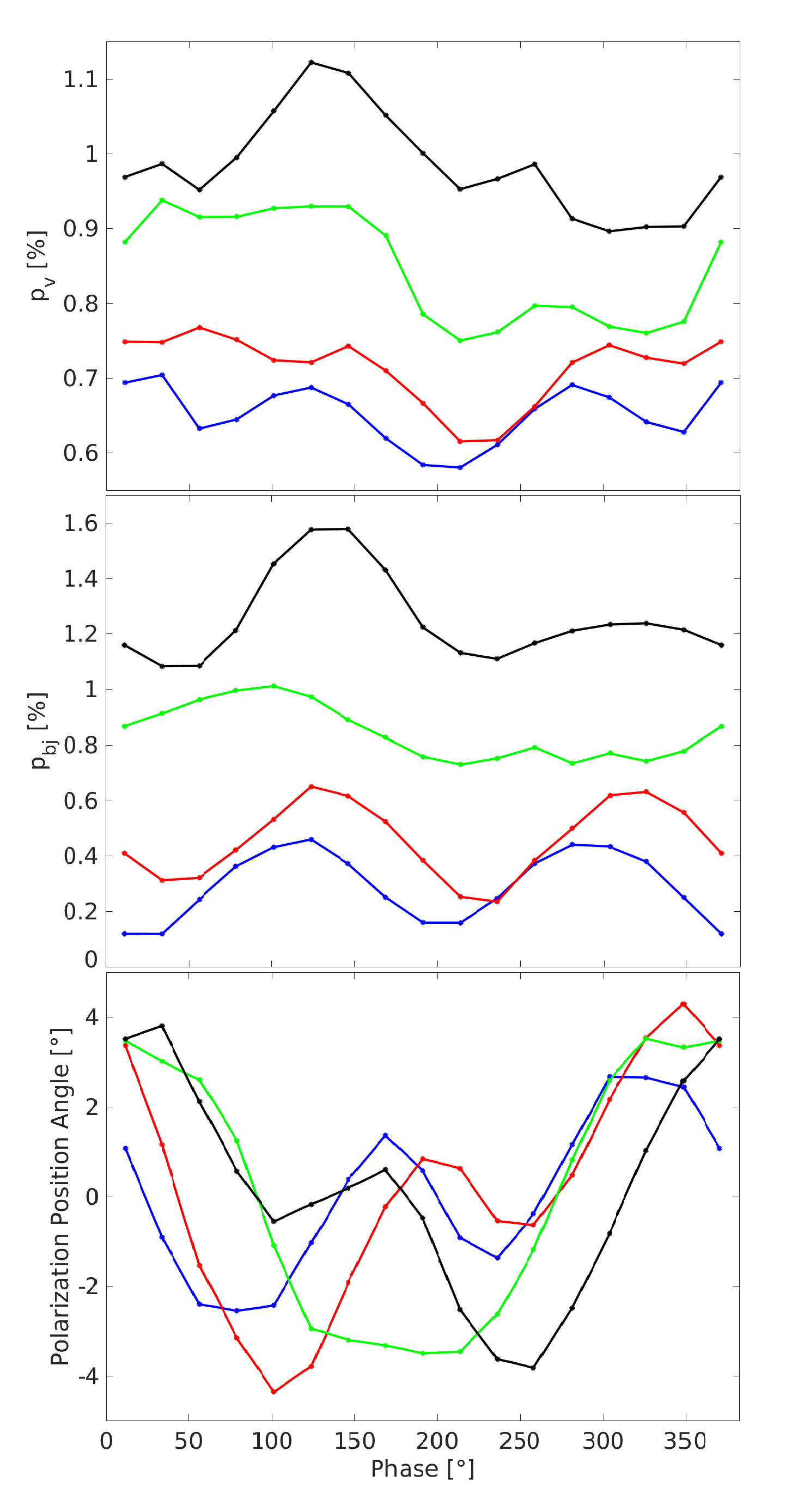}
\caption{Variation in percent polarization (top panel), Balmer jump polarization (middle panel), and PA (bottom panel) as a function of phase. The percent polarization and PA were measured in the V-band (data averaged from 5250 \AA{} and 5750 \AA). The Balmer jump polarization was computed as the difference in the continuum polarization at 3800 \AA{} and 3600 \AA{}. DE sections of the disk, $\omega$, were modelled to span 45$^{\circ}$, 90$^{\circ}$, 180$^{\circ}$ and 270$^{\circ}$, represented by blue, red, green and black respectively. The disk is viewed at $i$ = 70$^{\circ}$, with $n$ = 3.0, $\rho_0$ = 1.0 $\times$ 10$^{-11}$ g cm$^{-3}$ and an enhanced density of $\rho_0$ = 1.0 $\times$ 10$^{-10}$ g cm$^{-3}$.}
\label{fig:dem-pol}
\end{figure}

In the DE model, both the inclination of the system and the angular size of the DE sector must be considered when interpreting the polarization signature. First, the amount of polarization increases as the size of the DE sector increases. In other words, an overall denser disk contains more scattering material, which increases the amount of polarized light observed for disks with asymmetric polarization vectors. However, for optically thick regimes, the polarized level  may be diminished through absorption of the scattered photons. In addition to the size of the DE, the inclination of the system also affects the observed polarization. Figure \ref{fig:dem-pol} shows the polarization signatures of the DE model, for different sizes of DE as described in the figure caption.

In the top panel of Figure \ref{fig:dem-pol}, the V-band percent polarization increases as the size of the DE increases. In each case, the polarization is decreased as the enhanced density region is behind the star and then increases as it becomes visible again. For example, when $\omega = 270^{\circ}$ the polarization increases from around 0.95\% to 1.15\% as the lower density region is occulted. The opposite case, but same effect, is seen when $\omega = 90^{\circ}$. When $\omega = 270^{\circ}$, we see relatively strong polarization when the lower density region is behind the star which decreases as the higher density region moves behind the star.

The middle panel of Figure \ref{fig:dem-pol} shows how the size of the Balmer jump, measured in percent polarization, varies with phase. This was computed by taking the difference in the continuum polarization in 100 \AA{} bands on either side of the limit, centered roughly at 3600 \AA{} and 3800 \AA{}. When $\omega$ = $45^{\circ}$ and $90^{\circ}$, there is an increase in polarization received across first ${\sim}125^{\circ}$ of phase rotation. This occurs when the DE region is visible with the lower density parts of the disk either in front or behind it (i.e. when the DE region is viewed from the side). When looking through the DE region with the lower density region enveloping it, the scattering into the line of sight yields greater polarization as opposed to when the DE region lies directly in front of the star with respect to the observer.

\begin{figure}
\centering
\includegraphics[width=\columnwidth]{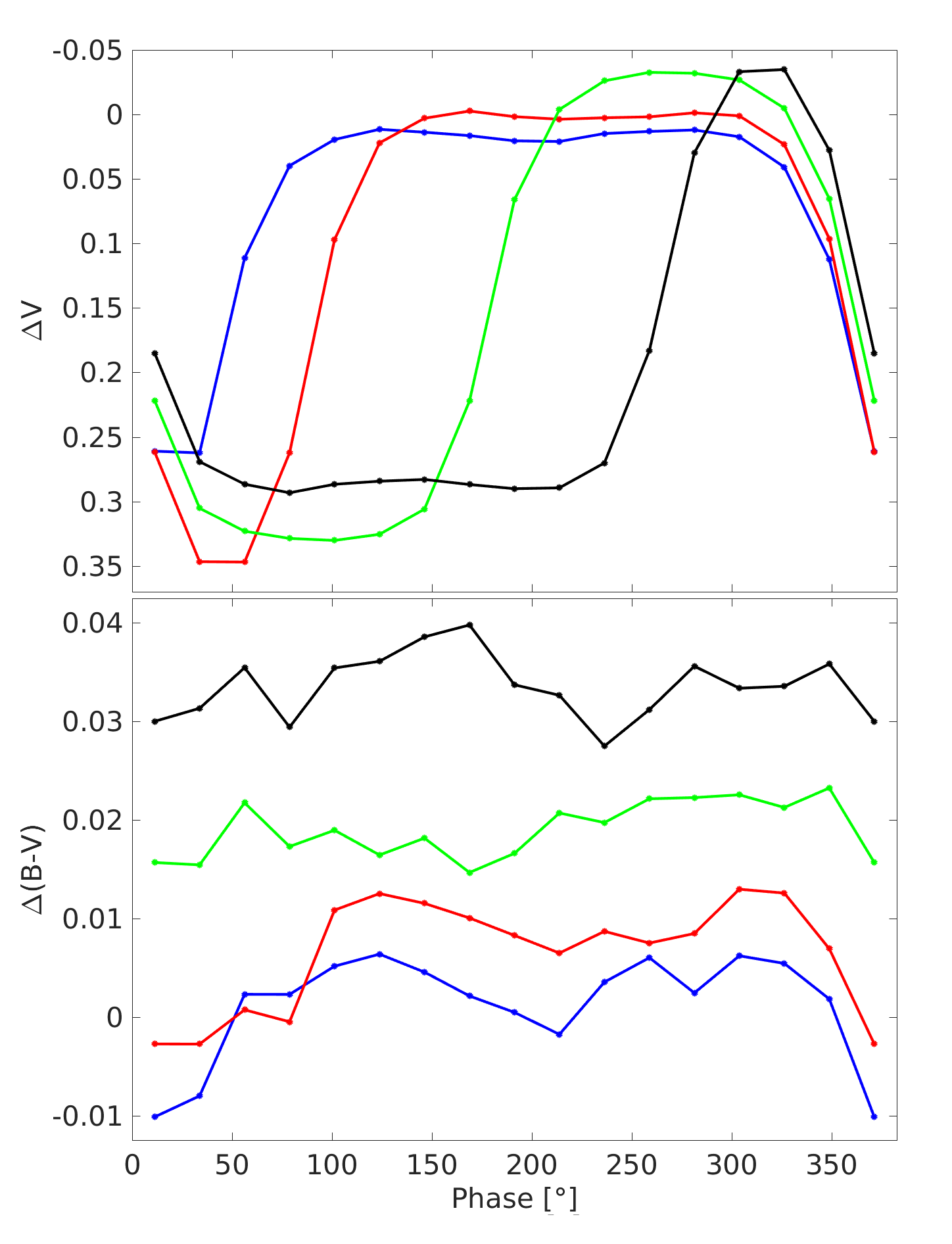}
\caption{$\Delta V$ (top panel), $\Delta (B - V)$ (bottom panel) colours as a function of phase. The figure colours and model parameters are the same as Figure \ref{fig:dem-pol}.}
\label{fig:dem-col}
\end{figure}

The bottom panel of Figure \ref{fig:dem-pol} shows the variation of PA with phase. The PA vector tracks the location of the DE, allowing for the position of the DE to be known (i.e. where the greatest source of polarization is). In each case, the PA oscillates around 0$^{\circ}$ as the disk rotates. For this system, the $u$ parameter varies from positive to negative (and is 0 when transitioning between) and dominates over the $q$ parameter, by virtue of how they were defined in \textsc{mctrace}. In the case where $\omega = 90^{\circ}$, the maximum PA is seen when viewing from phase = $0^{\circ}$. When the disk is rotated by $90^{\circ}$, the $u$ parameter becomes negative, and the minimum PA occurs. Rotating by another $90^{\circ}$ (i.e. phase = $180^{\circ}$), $u$ becomes positive once again, although because the DE sits partially behind the star, the maximum PA is reduced. The same effect occurs when phase = $270^{\circ}$, where the minimum PA is also reduced. The disk then rotates back to the starting position. This same behaviour is seen when $\omega = 45^{\circ}$, and in an inverse manner for $\omega = 270^{\circ}$. When $\omega = 180^{\circ}$ the $u$ and $q$ parameters all cancel out except for what has been occulted by the star, so the maximum PA occurs at phase = $0^{\circ}$, and the minimum at phase = $180^{\circ}$.

We used the Anderson-Darling (AD) and Cramér-von Mises (CvM) tests to demonstrate that our results come from different distributions. For the DE model these tests were applied to the results corresponding to $45^{\circ}$ and $90^{\circ}$, since our predictions for these warp sizes have the smallest observed difference. For the data presented in Figure \ref{fig:dem-pol} the largest p-values acquired were 8.0 $\times$ 10$^{-3}$ for the AD test, and 5.8 $\times$ 10$^{-3}$ for the CvM test.

\subsection{Colour-Magnitudes} \label{subsec:dem-mag}

\begin{figure}
\centering
\includegraphics[width=\columnwidth]{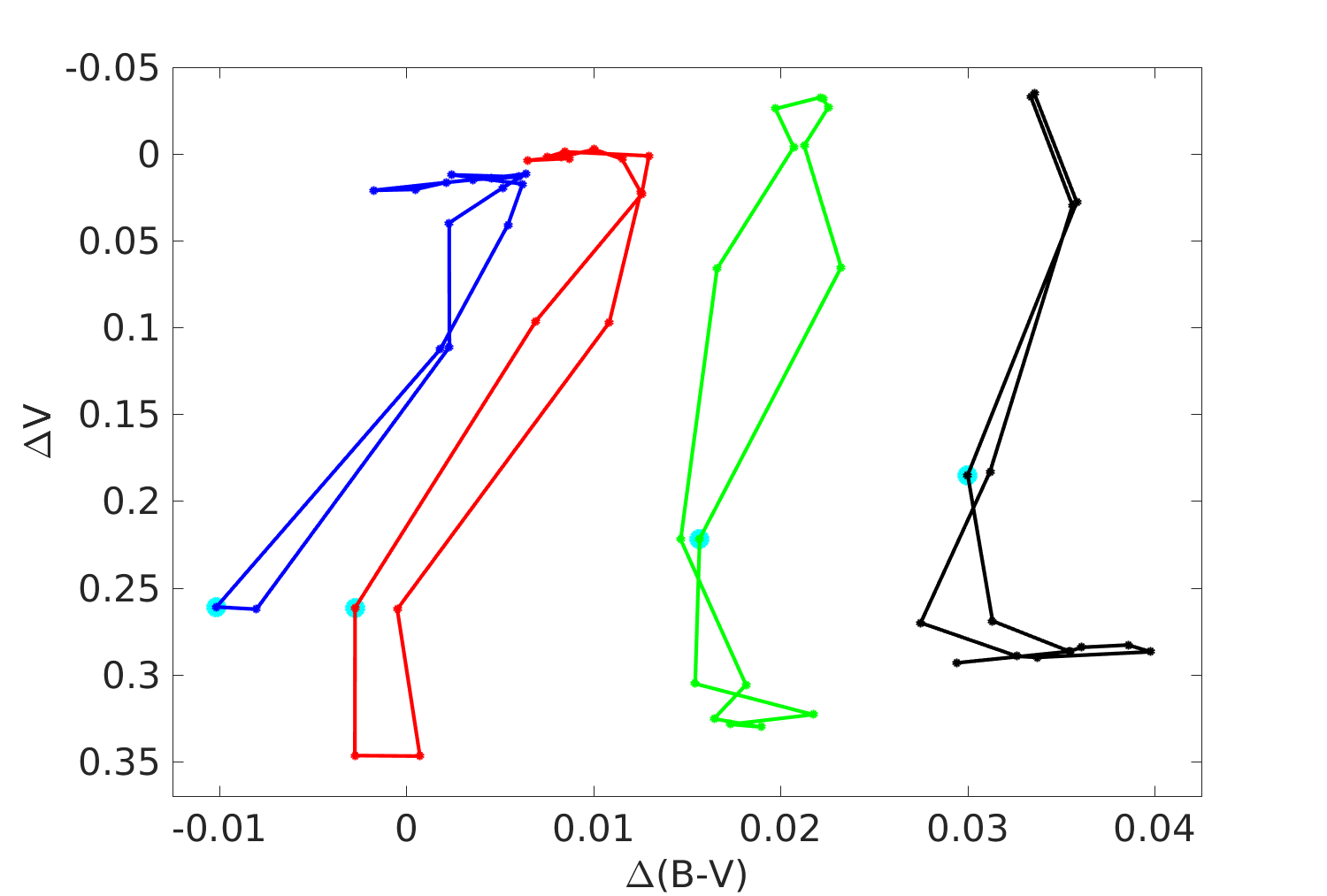}
\caption{$\Delta V$ versus $\Delta(B - V)$ magnitude-colour diagram with variation of phase. For each case, a cyan outline labels the data point corresponding to the first phase angle. The figure colours and model parameters are the same as Figure \ref{fig:dem-pol}.}
\label{fig:dem-vbv}
\end{figure}

The top panel of Figure \ref{fig:dem-col} shows how the V-band magnitude changes with phase. Overall the brightness is greatest when viewing from the low density region of the disk where the star may be seen. As the disk rotates to the high density region, the V-band magnitude decreases as the disk occults the star. The fall of the final data points in each set corresponds to the system becoming dimmer as the denser material returns into view and occults the star.

As the angular size of the DE sector increases, the system becomes redder as seen in Figure \ref{fig:dem-col}. In addition, the system becomes redder as the disk occults the star, although less so for larger DE angular size, making it challenging to discern any change when $\omega = 270^{\circ}$. When $\omega = 45^{\circ}$ and $90^{\circ}$, the effect of the DE going behind the star is also apparent in the data, as the DE moves behind the star making the system appear bluer.

The magnitude-colour loops shown in Figure \ref{fig:dem-vbv} combine the behaviours shown in both panels of Figure \ref{fig:dem-col}, summarized in the previous two paragraphs. The loops are diagonally oriented, appearing dimmer in the V-band while simultaneously bluer when viewing the system from a phase corresponding to a DE region. As the disk rotates and the system is seen from a phase corresponding to a lower density region, the system appears brighter and redder, as these parts of the disk have a smaller scale height, and occult less of the star. There is no apparent rotational direction to these loops (clockwise vs counter-clockwise), as the scale on which these observables vary is small for this particular model.

The AD and CvM tests were applied to the $\Delta V$ and $\Delta(B - V)$ results in Figures \ref{fig:dem-col} and \ref{fig:dem-vbv}, using the same method as used in Subsection \ref{subsec:dem-pol}. The resulting p-values from the AD test were virtually 0, which in this case means there is essentially no possibility that data sets tested originate from the same distribution. The largest p-value from the CvM test was found to be 4.4 $\times$ 10$^{-16}$.

\section{Tilted Disk Model} \label{sec:tdm}

A TD model was first proposed by \citet{hummel1998} in an effort to explain the quasi-cyclic variability of emission-line width that occurs over years and decades. His model suggested that a Be star's disk can be perturbed through an unknown mechanism, resulting in the disk being tilted from the equatorial plane. The formation of TD is commonly thought to involve tidal interaction between a binary companion and the disk e.g \citep{paploizou1995, martin2011, cyr2017}. Over the course of decades, the disk tilts to align with the companion's orbital plane, although \citet{cyr2017} found the disk tilts such that it is $90^{\circ}$ from the line of nodes of the companion's orbital plane. As the tilted disk rotates in the equatorial plane, it appears to an observer that the inclination of the Be star is rapidly changing. In this section, we investigate the effect a TD has on the system's polarization signature, magnitudes, and colours. These observables were analyzed at the same wavelengths as in the DE model.

\begin{figure}
\centering
\includegraphics[width=\columnwidth]{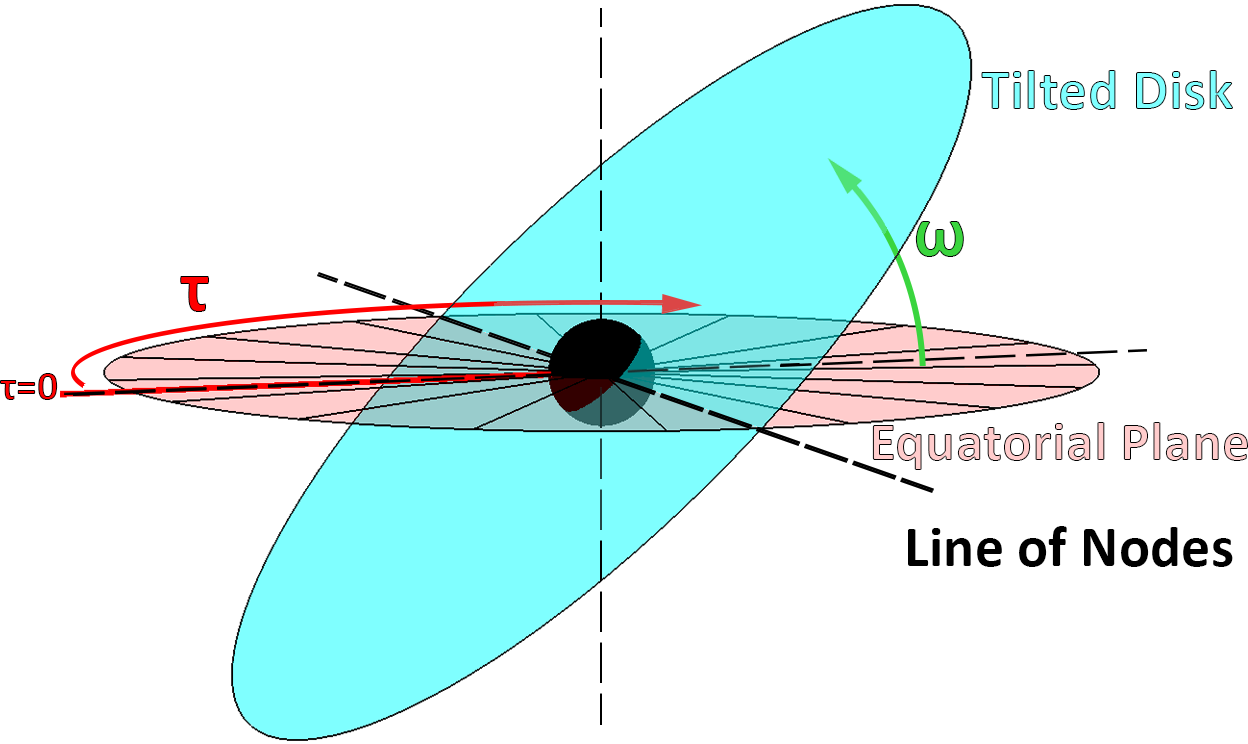}
\caption{Geometry of the tilted disk model showing the tilt angle, $\omega$, and the phase angle. The 16 wedges of the equatorial plane, on which the phase angles are centered, are shown in red.}
\label{fig:modeltilted}
\end{figure}

\textsc{bedisk} was used as a starting point for creating a reference model (see Tables \ref{tab:param1} and \ref{tab:param2} for the relevant parameters). \textsc{mctrace} was then used to calculate the Stokes parameters and photometric colours at 18 different inclinations and 16 phase angles (each phase angle corresponds with the middle of a disk sector, as shown in Figure \ref{fig:outfilepositions}). To emulate a TD, each phase angle was assigned a different inclination corresponding to the apparent inclination the Be star that an observer would see in the plane of the sky. This method creates the effect that as the system rotates in the equatorial plane, the Be star's inclination changes from the maximum value (equivalent to the tilt angle $\omega$) to edge-on with just a $90^{\circ}$ rotation of the disk, and to negative the maximum inclination (-$\omega$) as it rotates another $90^{\circ}$. Figure \ref{fig:modeltilted} illustrates the geometry of the TD model and the relevant variables associated with this work.

In order to determine the Stokes parameters of the TD, the Stokes parameters of the untilted disk were translated using the relevant inclination for the phase angle being considered. We use Equations \ref{eqn:uu} and \ref{eqn:qq} for each corresponding inclination of the wedge, where the untilted polarization p is in the desired wavelength band.

\begin{equation}\label{eqn:uu}
u_{tilted} = -p \cdot \sin(2\omega) \cdot sin(\text{\rm{PA}})
\end{equation}

\begin{equation}\label{eqn:qq}
q_{tilted} = \sqrt{p^2-u_{tilted}^2}
\end{equation}

Once the tilted Stokes parameters were obtained the amount of polarized light and the PA could be determined using Equations \ref{eq:pol} and \ref{eqn:polposangle}.

\subsection{Polarization} \label{subsec:tdm-pol}

\begin{figure}
\centering
\includegraphics[width=\columnwidth]{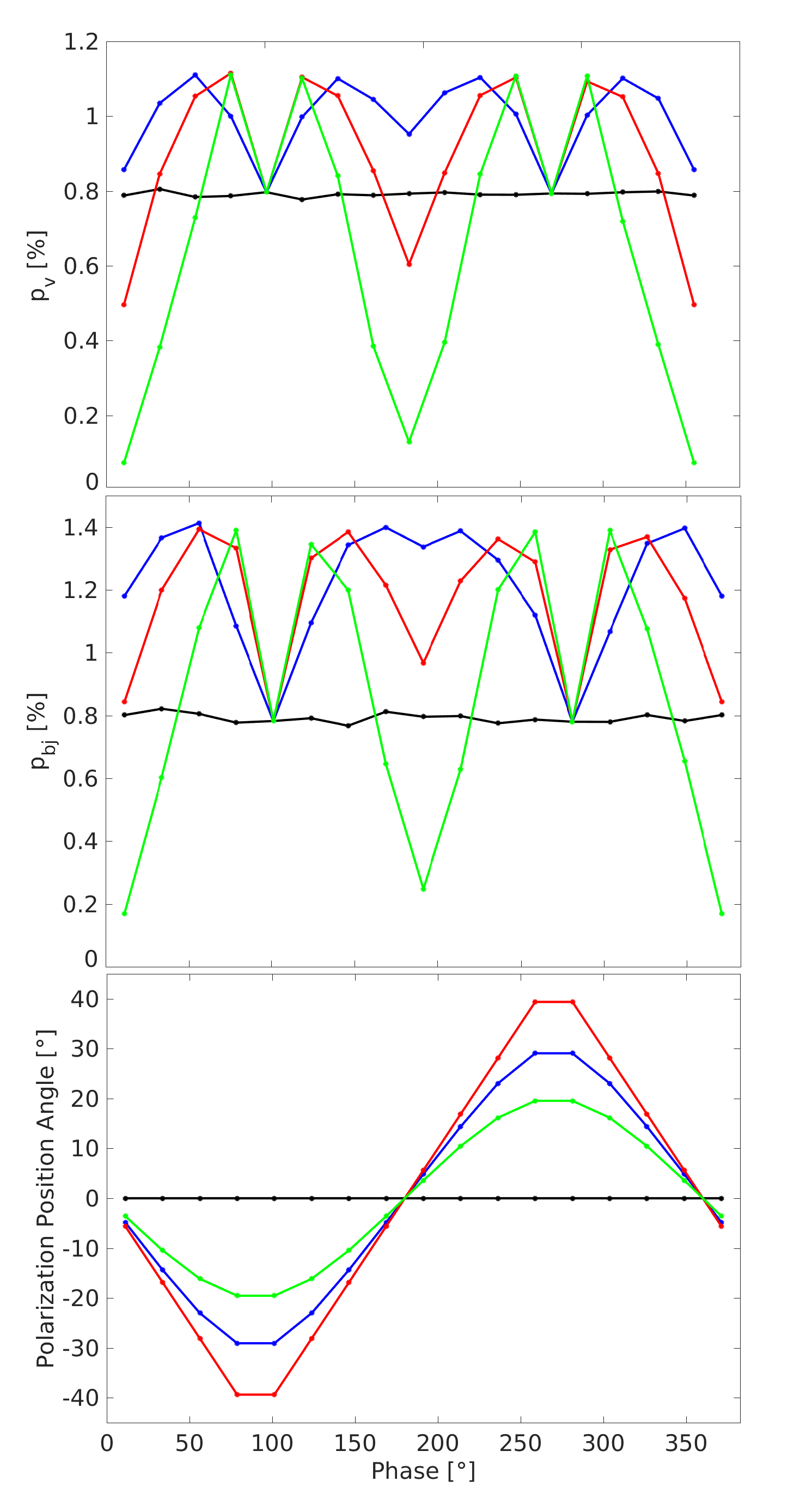}
\caption{Variation in percent polarization (top panel), Balmer jump polarization (middle panel), and PA (bottom panel) as a function of phase, for the TD model. The percent polarization and PA were measured in the V-band. The Balmer jump polarization was computed as the difference in the continuum polarization on either side of the limit, from $\lambda$ = 3800 \AA{} to 3600 \AA{}. Tilt angles, $\omega$, of $0^{\circ}$, $30^{\circ}$, $45^{\circ}$, and $70^{\circ}$  were used, represented by black, blue, red, and green respectively. The models were computed using the reference model parameters $n$ = 3.0 and $\rho_0$ = 5.0 $\times$ $10^{-11}$ g cm$^{-3}$, and is viewed from $i$ = $90^{\circ}$ (edge on to the equatorial plane).}
\label{fig:tdm-pol}
\end{figure}

The top panel in Figure \ref{fig:tdm-pol} shows the percent of polarized light as the disk undergoes a full rotation. When the system is not tilted (i.e. $\omega$ = $0^{\circ}$) the disk is consistently viewed from edge on, so no change in polarization is observed. As the tilt angle increases, the polarization minima (when the phase angle = $0^{\circ}$ and 180$^{\circ}$) continue to decrease as the symmetry of the system on the plane of the sky increases and further cancellation of orthogonal polarization vectors occurs. As the phase angle increases the disk rotates to be viewed edge on at phase angle = $90^{\circ}$ and $270^{\circ}$. At these phases, the percent polarization for a TD is equal to that of an untilted disk (i.e. $\omega$ = $0^{\circ}$) as the systems are geometrically equivalent. Noting that $\omega$ is measured from the equatorial plane, and as such is offset from inclination by $90^{\circ}$, the largest polarization when phase angle = $0^{\circ}$ is seen when $\omega$ = $30^{\circ}$, which is equivalent to viewing an untilted disk with i = $70^{\circ}$.

The middle panel of Figure \ref{fig:tdm-pol} shows how the Balmer jump polarization varies with phase. This was computed in the same manner as the middle panel of Figure \ref{fig:dem-pol}. As with Figure \ref{fig:dem-pol}, in the similarity between the Balmer jump polarization and the percent polarization at 3800 \AA{} (top panel) for the TD model occurs because the continuum at 3600 \AA{} is dominated by absorption and the fraction of polarized light at this wavelength is almost zero.

The PA as a function of phase angle for the TD is shown in the bottom panel of Figure \ref{fig:tdm-pol}. The greatest deviations from $0^{\circ}$ is seen when $\omega = 45^{\circ}$, beyond which the PA decreases as the disk tilts closer to pole on.

\begin{figure}
\centering
\includegraphics[width=\columnwidth]{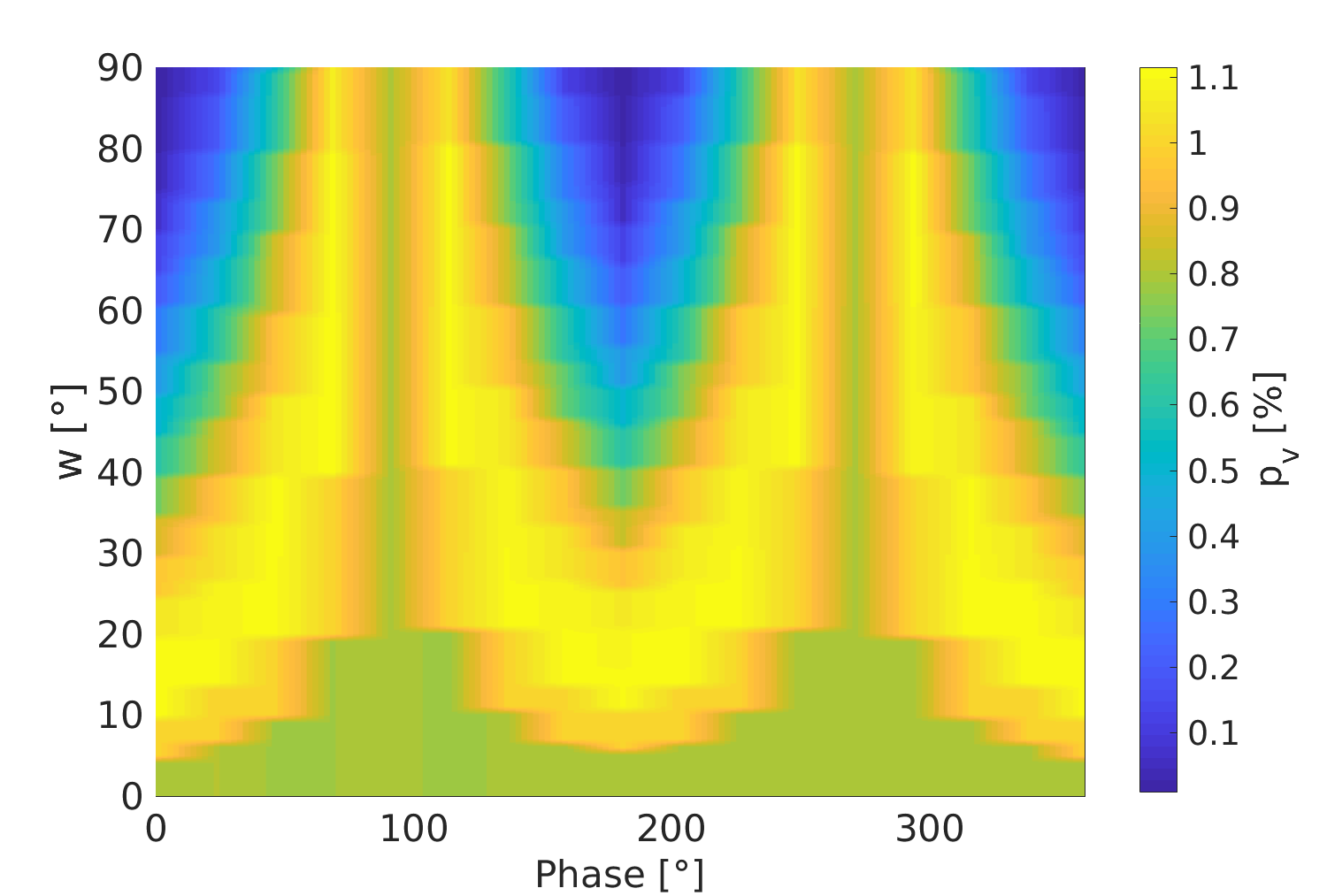}
\caption{Variation of percent polarization as $\omega$ and phase angle change. The same data as the top panel of Figure \ref{fig:tdm-pol} is shown, but in this heat map, the data is interpolated across all $\omega$ (from edge-on to pole on tilt angle) and phase angles (a complete rotation of the disk). The model parameters are the same as Figure \ref{fig:tdm-pol}.}
\label{fig:tdm-map}
\end{figure}

When considering a TD system, the maximum polarization signature is observed when $\omega \approx 30^{\circ}$ only when the phase angle $\approx 0^{\circ}$ and $180^{\circ}$. As the disk rotates, the maximum polarization will occur for different $\omega$'s at different phase angles. Figure \ref{fig:tdm-map} illustrates this phenomenon. Here we see that the variation of percent polarization becomes more significant for greater $\omega$, with relatively little change for small $\omega$, as expected.

The AD and CvM tests were applied to the polarization signatures in Figure \ref{fig:tdm-pol}, using the same method as used in Subsection \ref{subsec:dem-pol}. The largest p-values acquired are 2.1 $\times$ 10$^{-6}$ using the AD test, and 2.1 $\times$ 10$^{-3}$ using the CvM test.

\newpage

\subsection{Colour-Magnitudes} \label{subsec:tdm-mag}

\begin{figure}
\centering
\includegraphics[width=\columnwidth]{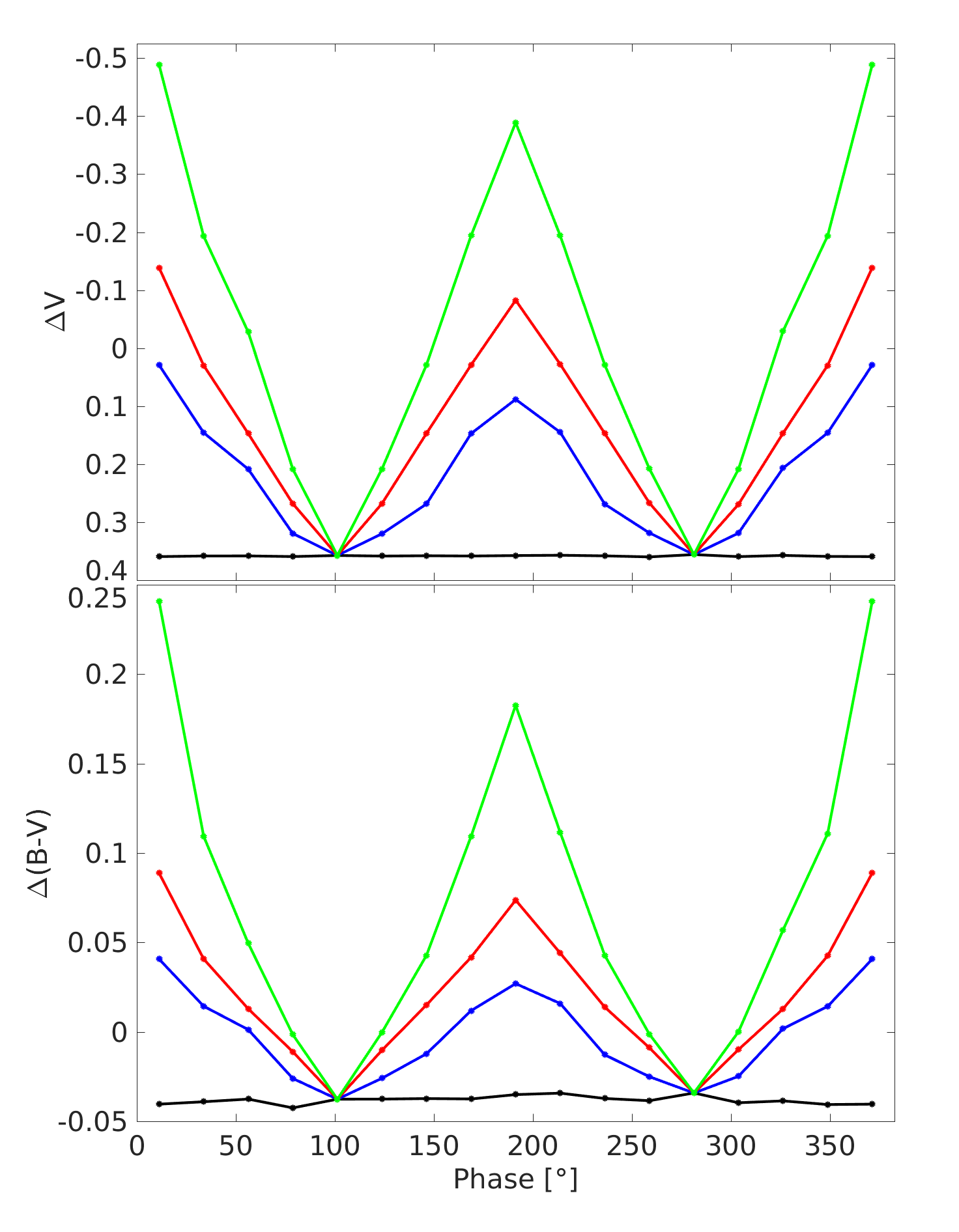}
\caption{$\Delta V$ (top panel) versus $\Delta(B-V)$ (bottom panel) colours as a function of phase. The figure colours and model parameters are the same as Figure \ref{fig:tdm-pol}.}
\label{fig:tdm-col}
\end{figure}

The top panel of Figure \ref{fig:tdm-col} shows the change in $\Delta$V as the disk rotates. As $\omega$ increases, the maximum brightness in the V-band (occuring at phase angle ${\sim}0^{\circ}$ and ${\sim}180^{\circ}$, when the disk is face-on) also increases. When the disk is edge on, the minimum brightness is seen, since the disk occults the star, in addition to a reduction of surface area of the disk.

The $\Delta(B-V)$ colour shown in the bottom panel of Figure \ref{fig:tdm-col} is inverted from the $\Delta$V, as the B-band light is significantly weaker than the V-band. As expected, the system becomes redder when the star is along the line of sight, and bluer when it becomes occulted, which may be affected by both the phase angle and the tilt angle.

\begin{figure}
\centering
\includegraphics[width=\columnwidth]{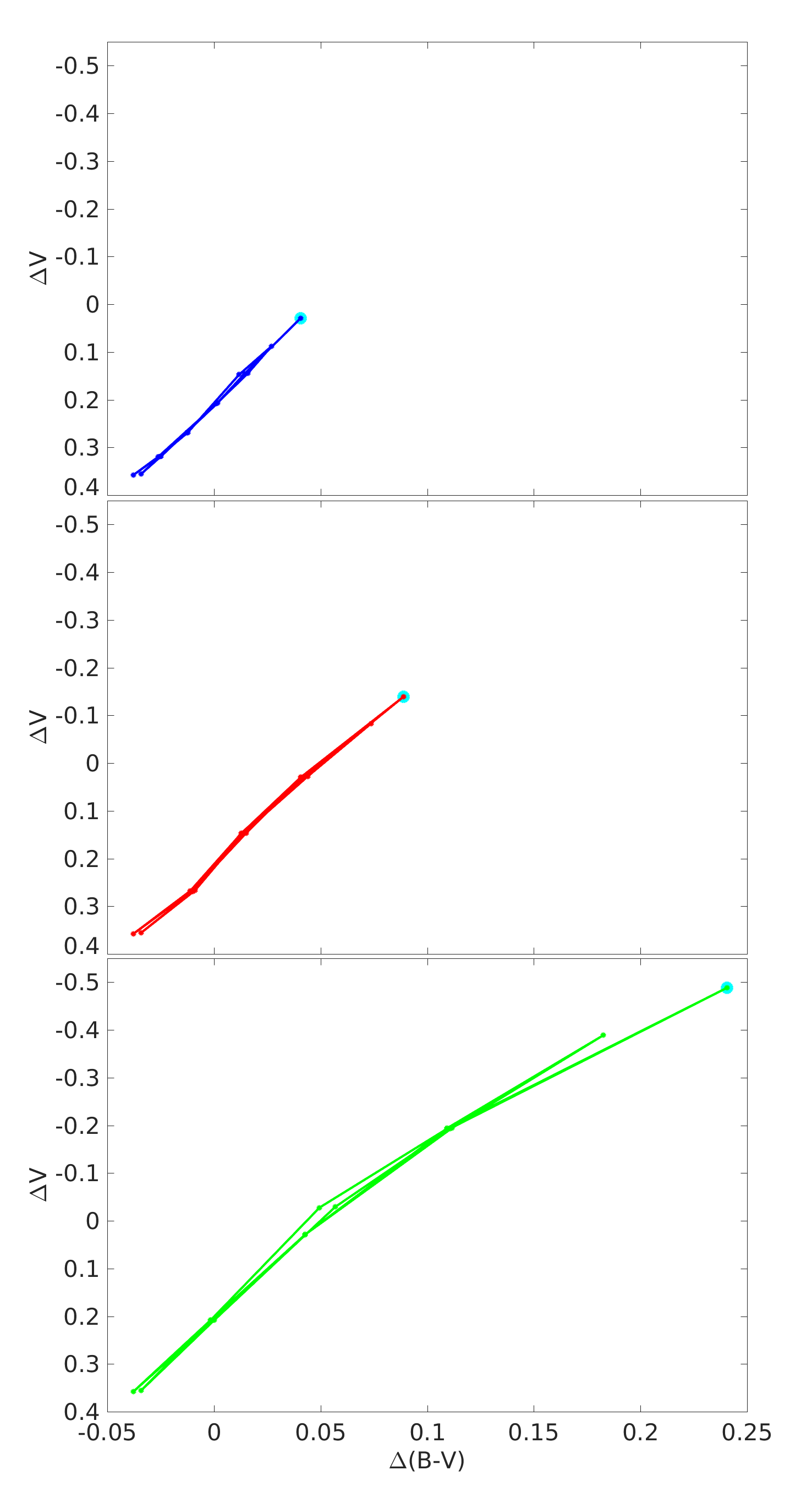}
\caption{$\Delta V$ versus $\Delta(B-V)$ as the phase goes from 0$^{\circ}$ to 360$^{\circ}$. Each simulation begins at the bottom right and with every complete revolution two loops are created. For each case, a cyan outline labels the data point corresponding to the first phase angle. The figure colours and model parameters are the same as Figure \ref{fig:tdm-pol}.}
\label{fig:tdm-colmag}
\end{figure}

The colour magnitude loops shown in Figure \ref{fig:tdm-colmag} combine the behaviours shown in both panels of Figure \ref{fig:tdm-col}, summarized in the previous two paragraphs. Similar to the DE model, we see the diagonal orientation of the loop, however the dimming in the TD model occurs when the rotation causes the disk to occult the star, also causing the system to appear bluer as star's V-band light dominates its B-band light.

We note that the variation in magnitude and colour would not be apparent for an axisymmetric system viewed edge-on, as the occultation of the star by the disk would not change over time. As such, the case of $\omega = 0^{\circ}$ was intentionally not shown in Figure \ref{fig:tdm-colmag}, since no sensible loop is observed.

The AD and CvM tests were applied to the $\Delta V$ and $\Delta(B - V)$ results in Figures \ref{fig:tdm-col} and \ref{fig:tdm-colmag}, using the same method as used in Subsection \ref{subsec:dem-pol}. The resulting p-values from the AD test were again virtually 0. The largest p-value from the CvM test was found to be 4.4 $\times$ 10$^{-16}$.

\section{Discussion and Summary} \label{subsec:disc}

As mentioned in Section \ref{sec:models}, the variability in the polarized light from a Be star may be used to study the geometry of the circumstellar material as the disk evolves. Values of percent polarization are between 0 to 1.5\%, with only ${\sim}5\%$ of Be stars lying above this range \citep{yudin2001}. \citet{wisniewski2010} showed that through the long-term process of disk growth and dissipation (on the order of decades), some stars can show a large degree of variation in polarization, with $\pi$ Aqr and 60 Cyg (HD 200310) varying up to 1\% or greater. These two stars in particular are interesting as both are thought to have binary companions and have recorded episodes of disk loss. Another Be star of interest is Pleione (HD 23862) which, according to \citet{tanaka2007}, has a double disk in which an old and TD is dissipating as a new disk is forming in the equatorial plane. In another study, \citet{hirata2007} presented an analysis on Pleione's TD and showed that the star's polarization has varied as much as ${\sim}0.5\%$ over time. \citet{draper2014} reported on the variable polarization of $\gamma$ Cas, and showed that it's V-band polarization had varied as much as 0.3\%. Be stars more often exhibit smaller variations in percent polarization rather than larger, as larger variations require more substantial changes to the disk as the observer sees it.

In the DE model, we find that the variability of the percent polarization, as seen in Figure \ref{fig:dem-pol}, is comparatively small, where the greatest change in the V-band is ${\sim}0.2\%$, the same maximum variability reported by \citet{huang1989}, and across the Balmer jump we see up to ${\sim}0.5\%$ variation. In contrast, Figure \ref{fig:tdm-pol} shows that the TD model also showed most values of V-band and Balmer jump polarization within ${\sim}0.5\%$, although they varied significantly more within that range than the DE model did. The greatest variation in V-band polarization seen in the TD model is ${\sim}1\%$, when tilt angle $\omega = 70^{\circ}$, which is similar to the values reported by \citet{wisniewski2010} for disk growth and dissipation. These cases likely produce similar variability as they are both examples of significant changes to the disk. We see that for TD's with tilt angles less than $45^{\circ}$ the variation remains $\le0.5\%$, similar to the values reported by \citet{hirata2007}.

The PA was found to be variable in four Be stars, $o$ And (HD 217675), $\gamma$ Cas, 88 Her (HD 162732), and $\kappa$ Dra (HD 109387), as reported by \citet{vince1995}, in which the variations were as large as $30^{\circ}$. $\gamma$ Cas, in particular, was observed to have PA variations up to $10^{\circ}$ (excluding the data point from 1976 as the authors labelled an outlier), in addition to showing percent polarization of up to 1\%, similar to Figure \ref{fig:dem-pol}'s $\omega = 90^{\circ}$ DE, which varies over ${\sim}8^{\circ}$. The other three stars in their work show greater PA variability than $\gamma$ Cas, of ${\sim}50^{\circ}$, $33^{\circ}$ and $20.6^{\circ}$ for $o$ And, 88 Her, and $\kappa$ Dra respectively, which are similar to the large PA variations seen in the TD model. The variability of polarization for $\gamma$ Cas which \citet{draper2014} reported on, showed PA variations as much as $10^{\circ}$. Significantly larger variations in PA may occur, such as that in \citet{hirata2007} where he shows Pleione's PA has changed by ${\sim}80^\circ$ over the course of ${\sim}20$ years. The PA seen from the DE model is consistent with the values reported by Hirata, as shown in Figure \ref{fig:dem-pol}'s bottom panel. As with the percent polarization, the PA in the TD model changes much more significantly than in the DE model. The bottom panel of Figure \ref{fig:tdm-pol} shows that the largest possible variation in PA is ${\sim}80^\circ$ when the tilt angle is $45^{\circ}$, which is consistent with \citet{hirata2007}'s data on Pleione.

As previously mentioned, when comparing the TD and DE models, it is best to compare the DE model to the $\omega = 30^{\circ}$ TD model. This is because the DE model is inclined at $i = 70^{\circ}$ while the TD model's tilt angle is measured from the equatorial plane, making $\omega = 30^{\circ}$ in the TD model equivalent to $i = 70^{\circ}$ in the DE model.

Both models show variability of $\Delta$V and $\Delta(B-V)$ within one magnitude, which is consistent with the values of a group of 500 galactic Be stars reported by \citet{bartz2017}. Unlike the polarization signatures, the variation of the colour and magnitude are similar when comparing the $\omega = 30^{\circ}$ TD model with the DE model. The largest variations we see for the DE model in $\Delta$V are ${\sim}0.4$ magnitude, and in $\Delta$(B-V) are ${\sim}0.01$ magnitude, where all sizes of the DE show variations of approximately the same amount. For the TD model, the variations in the colour-magnitudes get larger as the tilt angle increases, as more of the star is visible. For the largest adopted tilt angle of $70^{\circ}$, we see variations in $\Delta$V of ${\sim}1$ magnitude, while the $\Delta$(B-V) varies ${\sim}1.2$ magnitudes.

\citet{harmanec1983} reports on the variability of Be stars, stating that when the V band becomes brighter the B-V index is expected to become redder, and U-B index is expected to become bluer, which he attributes to a positive correlation between H{\sc i} emission strength and the stellar luminosity, existing for long-term variations. Conversely, he also describes an inverse correlation between H{\sc i} emission strength and the stellar luminosity, such that as the V band becomes dimmer, both the B-V and U-B indicies become redder. In both the DE and TD models, we see the evidence of the positive correlation described by \citet{harmanec1983}'s, as the loops that result from the rotation of the disk in Figures \ref{fig:dem-vbv} and \ref{fig:tdm-colmag} are all positively sloped.

The same orientation and looping behaviour has been observed in the work of \citet{dewit2006} and \citet{jones2013}. \citet{dewit2006} studied the photometric variability in a sample of Be stars in the Small Magellanic Cloud, and found that ${\sim}90\%$ of the the stars had colour-magnitude loops that rotated in a clockwise sense, with the loops of the remainder stars rotating counter-clockwise. They found that an outflowing disk and variable mass loss could recreate the rotational behaviour of the loops. \citet{jones2013} showed the positive correlation behaviour described by \citep{harmanec1983} exists for $\delta$ Sco (HD 143275) for disk growth and dissipation events. They also show that $\delta$ Sco's loops rotate in a clockwise fashion, and vary less in both colour and magnitude than the TD model with a tilt angle of $30^{\circ}$ as shown in the top panel of Figure \ref{fig:tdm-colmag}.

The same behaviour again has been reproduced in models of Be stars, such as the work of \citet{haubois2012} and \citet{granada2017}. The disk growth and dissipation models reported in \citet{haubois2012}, with $\rho_0 = 3 \times 10^{-11}$ gcm$^{-3}$, which is an approximate average of the two densities adopted here for the DE model, produce similar loops of an equivalent scale as those presented in this work. In terms of average disk density, \citet{haubois2012}'s system is most comparable to the $\omega = 45^{\circ}$ DE model. When viewing at $i = 70^{\circ}$, \citet{haubois2012}'s model shows modest variations centered around $\Delta V = 0$ and $\Delta(B-V) = 0$, similar to what is seen in the $\omega = 45^{\circ}$ DE model when the enhancement is behind the star. Once the DE region becomes visible again, the brightness decreases as the DE region begins to once again occult the star. There is a small difference in $\Delta$V brightness between our model and \citet{haubois2012}'s, which may be explained by the greater average disk density adopted in our model. 

\citet{granada2017} presents theoretical colour-magnitude loops of Be stars in infrared wavelengths. It is important to note that the primary source of the infrared wavelengths is the disk as opposed to the V band wavelengths from the generated primarily by the star as seen in Fig. 2 of \citep{rivinius2013}. \citet{granada2017} showed that for Be stars with small and intermediate inclination angles that stars with stable and developed disks have $n\leq3.5$, and intermediate values of base density $\rho_0$, although for larger inclination angles the IR excess diminishes, as expected as less surface area of the disk is visible. In comparison with our TD models, where a large tilt angle is analogous to a small inclination, and vice versa, we see that as the tilt angle increases the variation of the V-band magnitude increases. So, as IR excess becomes smaller in \citet{granada2017}'s work, an equivalent change in a Be star in our work predicts smaller variation in V-band magnitude.

The hydrodynamic SPH simulations in \citep{cyr2017} showed that when the orbital plane of a binary companion is misaligned from the equatorial plane of the Be star (misalignment angle), the greatest tilting effect is seen at $45^{\circ}$, and the effect of tilting the disk diminishes from greater misalignment angles. Their results also showed that the tilt angle never becomes as large as the misalignment angle, and is generally less than half the misalignment angle. This in part may explain why the tilt angle = $30^{\circ}$ TD model, is consistent with with observations presented in the literature. 

In the DE model, all of the ranges of variability of the polarization and colour-magnitudes reported in this work are equally present in the literature. Hence, based on the models presented here, that we cannot places bounds on the size of the DE region.

In summary, by using polarimetry, important properties of the disks may be investigated, such as the geometry, chemistry, density, and disk evolution, without the need of resolving the disk. In this work, we created two models with different disk shapes, and used the multi-wavelength polarimetric and photometric signatures to analyze their disk behaviour. In future, we plan to merge the the DE and TD models to create a tilted disk with a DE region and expand our code to predict other key observables.

\acknowledgments
CEJ wishes to acknowledge support though the Natural Sciences and Engineering Research Council of Canada. The authors would like to thank Professor Roge Mamon from the Department of Statistical and Actuarial Sciences at The University of Western Ontario for helpful discussions related to the statistical analysis of our results.

\end{document}